\documentclass[superscriptaddress, amsmath,amssymb,
 aps, pra, twocolumn]{revtex4-2}
\usepackage{txfonts}

\usepackage{amsthm}

\usepackage{graphicx}% Include figure files
\usepackage{dcolumn}% Align table columns on decimal point
\usepackage{bm}% bold math
\usepackage{xcolor}
 
\usepackage{hyperref}% add hypertext capabilities
\hypersetup{colorlinks=true, citecolor=blue, linkcolor=blue}
\label{sec:convergence}

\theoremstyle{definition}

\newcommand{\bra}[1]{\langle #1 |}
\newcommand{\ket}[1]{| #1 \rangle}

\begin{document}

\title{Scalable linearized gate set tomography}
\author{Ashe Miller}
\affiliation{Quantum Performance Laboratory, Sandia National Laboratories, Livermore, CA 94550}
\author{Corey Ostrove}
\affiliation{Quantum Performance Laboratory, Sandia National Laboratories, Albuquerque, NM 87185}
\author{Jordan Hines}
\affiliation{Quantum Performance Laboratory, Sandia National Laboratories, Albuquerque, NM 87185}
\author{Noah Siekierski}
\affiliation{Quantum Performance Laboratory, Sandia National Laboratories, Livermore, CA 94550}
\author{Kevin Young}
\affiliation{Quantum Performance Laboratory, Sandia National Laboratories, Livermore, CA 94550}
\author{Robin Blume-Kohout}
\affiliation{Quantum Performance Laboratory, Sandia National Laboratories, Albuquerque, NM 87185}
\author{Timothy Proctor}
\affiliation{Quantum Performance Laboratory, Sandia National Laboratories, Livermore, CA 94550}
\date{September 2023}
\date{\today}

\begin{abstract}
Characterizing errors on many-qubit quantum computers remains a key challenge to understanding and improving the performance of these devices. Current characterization methods either don't scale beyond a few qubits, or make simplifying assumptions (such as assuming stochastic Pauli error) that obscure the underlying physical error mechanisms. In this work, we present a scalable extension to gate set tomography---\emph{linearized gate set tomography}---that enables characterization of many-qubit systems. Linearized gate set tomography relies on sparse error models, a linear approximation to enable efficient data fitting, and data from shallow circuits---so that the systematic error in the linear approximation is small. We demonstrate the accuracy of our technique using simulations of a ten-qubit system with coherent and stochastic errors, including coherent crosstalk, and we demonstrate that it is robust in presence of additional errors that are not included within the sparse error model ansatz. 
\end{abstract}

\maketitle
\section{Introduction}
Recent advances in quantum computing have resulted in quantum processors with hundreds of physical qubits \cite{wurtz2023aquilaqueras256qubitneutralatom, google2025quantum, IBM, ransford2025helios98qubittrappedionquantum, Chen2024benchmarkingtrapped}. However, these systems are still far from achieving the scale and error rates needed to achieve utility-scale fault-tolerant quantum computing \cite{benchmarking_perspective,gidney2021factor, PRXQuantum.2.030305, rubin2024quantum}. One of the major blockades to further improvements is the ability to understand the errors and noise that afflict these systems. State-of-the-art techniques for studying the errors in many-qubit quantum computers achieve scalability by measuring a small number of performance metrics, such as mean gate fidelity, or by enforcing and then learning simplified error models like Pauli noise \cite{magesan2011scalable, RBReview, eisert2020quantum, childs2018toward,PhysRevLett.123.030503, hines2024fully, mckay2023benchmarkingquantumprocessorperformance,PhysRevLett.129.150502, PhysRevX.13.041030, helsen2019new, PRXQuantum.3.020357, PhysRevLett.109.080505}. By design, these approaches are insensitive to many of the details about what types of errors are limiting performance making it difficult to use this information to improve or predict performance. Extracting detailed information about the errors that are occuring requires detailed characterization methods such as gate set tomography (GST) \cite{gatesettomographyReview2021,RLGST2021,cGST}. 

GST is a method for learning all of the Markovian errors present in each of a quantum computer's gates and its state preparation and measurement (SPAM). Unlike process tomography, GST is robust to SPAM errors and it enables high-precision estimation of the rates of every kind of Markovian error \cite{gatesettomographyReview2021}. GST on $n$ qubits learns an $n$-qubit gate set, where here a ``gate'' means an operation acting on all $n$ qubits (so, for $n > 1$ a ``gate'' often consists of parallel applications of one- and two-qubit gates, which in other contexts is instead typically called a ``layer'' or ``cycle''). Each gate is represented by an $16^n$-element process matrix with $\mathcal{O}(16^n)$ a priori unknown elements. This means that the experimental and computational cost of GST scales \emph{exponentially} in $n$. Even with optimized GST experiments \cite{rudinger2023twoqubitgatesettomography, Ostrove2023-zc}, GST is infeasible for more than approximately three qubits, and only one- and two-qubit GST experiments are efficient enough to be regularly performed.

The fundamental roadblock to scalable GST is that the number of parameters in its model scales exponentially---there are $\mathcal{O}(16^n)$ parameters for every gate. This additionally causes an exponential overhead when doing data post processing due to the exponential scaling it induces in the necessary exact simulation. Existing approaches to improving the efficiency of GST \cite{Ostrove2023-zc, rudinger2023twoqubitgatesettomography, cGST, RLGST2021} have decreased the cost of the standard GST protocol \cite{gatesettomographyReview2021} but none have addressed this foundational issue. Compressive gate set tomography (cGST) \cite{cGST}, GST experiment design reduction methods \cite{rudinger2023twoqubitgatesettomography, Ostrove2023-zc}, and randomized linear gate set tomography (RLGST) \cite{RLGST2021} reduce the number of circuits that need to be run but do not change the scaling with $n$. RLGST used an important insight: by only using data from shallow circuits it is possible to use a linear approximation that forgoes the computational cost of exact simulation of long amplification circuits, reducing the cost of modeling fitting and experiment design. But, RLGST still fits an error model with $\mathcal{O}(16^n)$ unknowns, making it unscalable.

In this paper, we introduce a variation of GST---\emph{linearized GST}---that combines sparse error models and approximate simulation methods \cite{ErrorPropagation} to enable fully scalable characterization of $n$-qubit models. Like RLGST, linearized GST characterizes gates using short circuits, and in our numerical experiments those circuits are also random. However, we achieve polynomial classical computational cost by utilizing the sparse error models and efficient approximate simulation techniques introduced in Ref.~\cite{ErrorPropagation}. We learn a polynomial number of unknowns, allowing for the characterization of many-qubit systems---which we demonstrate with simulation on up to ten qubits. 

In section~\ref{sec:SectionII} we review the small Markovian error generators on which our sparse error models are based. In section \ref{sec:SectionIII} we introduce our sparse error models, we show how learning these sparse error models can be approximated as a linear inversion problem, and we present the linearized GST protocol. In section \ref{sec:SectionIV}, we demonstrate linearized GST in simulation on up to ten qubits. We conclude in section \ref{sec:conclusions}.

\section{Error Generators}
\label{sec:SectionII}
\subsection{Small Markovian error generators}
We begin by reviewing the formalism behind our error models: elementary error generators \cite{SmallMarkovianErrors}. Linearized GST aims to characterize post-layer \cite{SmallMarkovianErrors} errors in an $n$-qubit quantum computer. Thus every layer of gates is modeled as the perfect unitary acting on the $n$ qubits followed by an error channel. We represent the action of a noisy layer of gates by
\begin{align}
    \tilde{\mathcal{U}} =e^{\mathcal{L}} \mathcal{U},
\end{align}
where $\mathcal{U}$ is the $n$-qubit superoperator for an ideal gate operation ($\mathcal{U}[\rho] = U\rho U^{\dagger}$ where $U$ is the unitary for the layer), and $\mathcal{L}$ is the post-layer error Lindbladian. We decompose the Lindbladian into the basis of elementary error generators \cite{SmallMarkovianErrors,ErrorPropagation}, given as
\begin{align*}
    \mathbb{H}_P[\rho]&=-i\left[P,\rho\right],\notag\\
    \mathbb{S}_P[\rho]&=P\rho P-\rho,\notag\\
    \mathbb{C}_{P,Q}[\rho]&=P\rho Q+Q\rho P-\{\{P,Q\},\rho\},\notag\\
    \mathbb{A}_{P,Q}[\rho]&=i\left(P\rho Q-Q \rho P +\{[P,Q],\rho\}\right),
\end{align*}
where $P,Q \in \mathbb{P}^*$ where $\mathbb{P}^*$ is the set of $n$-qubit Pauli operators without signs and excluding the identity, and
$Q > P$ according to some indexing of $\mathbb{P}$.  That is, we write $\mathcal{L}$ as 
\begin{align*}
   \mathcal{L} =\sum_{P\in\mathbb{P}^*} \left(h_{P} \mathbb{H}_P+s_{P} \mathbb{S}_P \right)+\sum_{Q, P \in\mathbb{P}^*, Q > P} \left(c_{P,Q}\mathbb{C}_{P,Q}+a_{P,Q} \mathbb{A}_{P,Q}\right),
\end{align*}
where the lower-case letters are the error rates for each type of error. Learning the error rates is equivalent to learning the Lindbladian.

For the rest of this paper, we will consider Lindbladians that can be expressed as a linear combination of $\mathbb{H}_P$ and $\mathbb{S}_P$ errors, i.e., we assume  $c_{P,Q} = a_{P,Q} = 0$. The $\mathbb{H}_P$ and $\mathbb{S}_P$ terms enable modeling of coherent and Pauli stochastic errors, respectively, which are two of the most common kinds of error in physical systems. We note, however, that most of the conceptual underpinnings of linearized GST can be applied to models also containing $\mathbb{C}$ and $\mathbb{A}$ terms.
The error map $e^{\mathcal{L}}$ is not completely positive and trace preserving (CPTP) for all possible values for the error rates---but the CPTP constraint is simple to enforce with only $\mathbb{H}$ and $\mathbb{S}$ errors. The error map is CPTP for any (real) values of the Hamiltonian error rates and for all non-negative values for the Pauli stochastic error rates \cite{SmallMarkovianErrors}.

\subsection{Clifford propagation of small Markovian error generators}
The elementary error generators basis can be leveraged for efficient approximate simulation of noisy Clifford circuits \cite{ErrorPropagation}. One of the key reasons for this is that Clifford unitaries preserve the basis of elementary error generators. That is, conjugation of any elementary error generator by the superoperator $\mathcal{U}$---where $\mathcal{U}[\rho]=U\rho U^{\dagger}$ with $U$ a Clifford unitary---transforms that elementary error generator into another elementary error generator, sometimes multiplied by a sign \cite{ErrorPropagation}. Therefore, if we have an error channel, written as the exponential of a Lindbladian represented in the elementary error generator basis, occurring before a Clifford operation, we can rewrite this as
\begin{align}
\label{Eqn:CliffordProp}
    &\mathcal{U}\exp\left[\sum_{P\in\mathbb{P}_{s}} h_{P}\mathbb{H}_P+s_{P}\mathbb{S}_P\right]\notag\\
    &\rightarrow\exp\left[\sum_{P\in\mathbb{P}_{\ell}}h_{P}\gamma\left(U,P\right)\mathbb{H}_{P'}+s_{P}\mathbb{S}_{P'}\right]\mathcal{U},
\end{align}
where $\mathbb{P}_{s}$ denotes the subset of Pauli operators that index the errors that occur in elementary error generator decomposition of this error channel, $P'=UPU^{\dag}$ modulo sign, and $\gamma(U,P) = \pm 1$ is also efficient to compute \cite{ErrorPropagation}.

To simulate a noisy Clifford circuit we can repeatedly apply Eqn.~\eqref{Eqn:CliffordProp} to propagate all of the error maps---i.e., an error map for each layer in the circuit---to the end of the circuit. For the rest of this paper, the resulting product of error channels will be termed the \emph{post-circuit error channel}. We can write a general post-circuit error channel for a circuit with layers $\ell$ as
\begin{align}
\label{Eqn:ErrorProp}
\varepsilon=\prod_{\ell}\exp\left[\sum_{P\in\mathbb{P}_{\ell,H}}h^{(\ell)}_{P}\gamma\left(U_{:\ell},P\right)\mathbb{H}_{P'}+\sum_{P\in\mathbb{P}_{\ell,S}}s^{(\ell)}_{P}\mathbb{S}_{P'}\right],
\end{align}
where $\mathbb{P}_{\ell,H}$ and $\mathbb{P}_{\ell,S}$ denotes the set of Pauli operators that index the Hamiltonian and Pauli stochastic errors, respectively, that occur in layer $\ell$, with corresponding errors $h_{P}^{(\ell)}$ and $s_{P}^{(\ell)}$, and $P' = U_{:\ell} P U_{:\ell}^{\dag}$ with $U_{:\ell}$ denoting the composition of all the unitaries for all the layers occurring after layer $\ell$.

Circuits with a relatively small amount of error can be accurately approximated by the \emph{first-order post-circuit error channel} \cite{ErrorPropagation}:
\begin{align}
\varepsilon\approx\mathbb{I}+\sum_{\ell}\left(\sum_{P\in\mathbb{P}_{\ell,H}}h^{(\ell)}_{P}\gamma\left(U_{:\ell},P\right)\mathbb{H}_{P'}+ \sum_{P\in\mathbb{P}_{\ell,S}}s^{(\ell)}_{P}\mathbb{S}_{P'}\right),
\label{eq:circuit-first-order}
\end{align}
which is obtained from a Taylor expansion of Eqn.~\eqref{Eqn:ErrorProp}. The error in this approximation is order $O(h^2,s^2)$ where $h$ and $s$ are constants such that $h > |h_{\ell,P}|$ and $s > s_{\ell,P}$ for all $\ell$ and $P$ \cite{ErrorPropagation}. From the first-order post-circuit error channel, we can efficiently calculate measurement outcome probabilities and expectation values. In this work, we will be using Pauli operators measurements, and to compute how $\varepsilon$ impacts such measurements we can compute the impact of each of the terms in Eq.~\eqref{eq:circuit-first-order}. The effect of a stochastic error generator $\mathbb{S}_Q$ on a Pauli observable $P$ is
\begin{align}
\label{eqn:Ssen}
\text{Tr}\left[P\mathbb{S}_Q\left(\ket{\psi}\bra{\psi}\right)\right]=\begin{cases}
        \pm 2 & \left[P,Q\right]\neq 0 \text{ and } P\ket{\psi}=\mp \ket{\psi}\\
        0 & \text{otherwise},
    \end{cases}
\end{align}
where $\ket{\psi}$ is the (stabilizer) state produced by the circuit in the absence of errors. Similarly, the impact of a  coherent error $\mathbb{H}_Q$ on the Pauli observable $P$ is
\begin{align}
    \label{eqn:Hsen}
    \text{Tr}\left[P\mathbb{H}_Q\left(\ket{\psi}\bra{\psi}\right)\right]=\begin{cases}
        \pm 2 & \left[P,Q\right]\neq 0 \text{ and } PQ\ket{\psi}=\pm \ket{\psi} \\
        0 & \text{otherwise}.
    \end{cases}
\end{align}
These equations imply that for any specific circuit, each Pauli observable will be affected to leading order \textit{only} either by $\mathbb{H}$ or by $\mathbb{S}$ errors, but not both. Eqn.~\eqref{eqn:Ssen} and Eqn.~\eqref{eqn:Hsen}, and all of the error propagation formulae, can be efficiently evaluated \cite{ErrorPropagation}.

\section{Theory for Scalable Sparse Error Learning}
\begin{figure*}
    \centering
    \includegraphics[width=15cm]{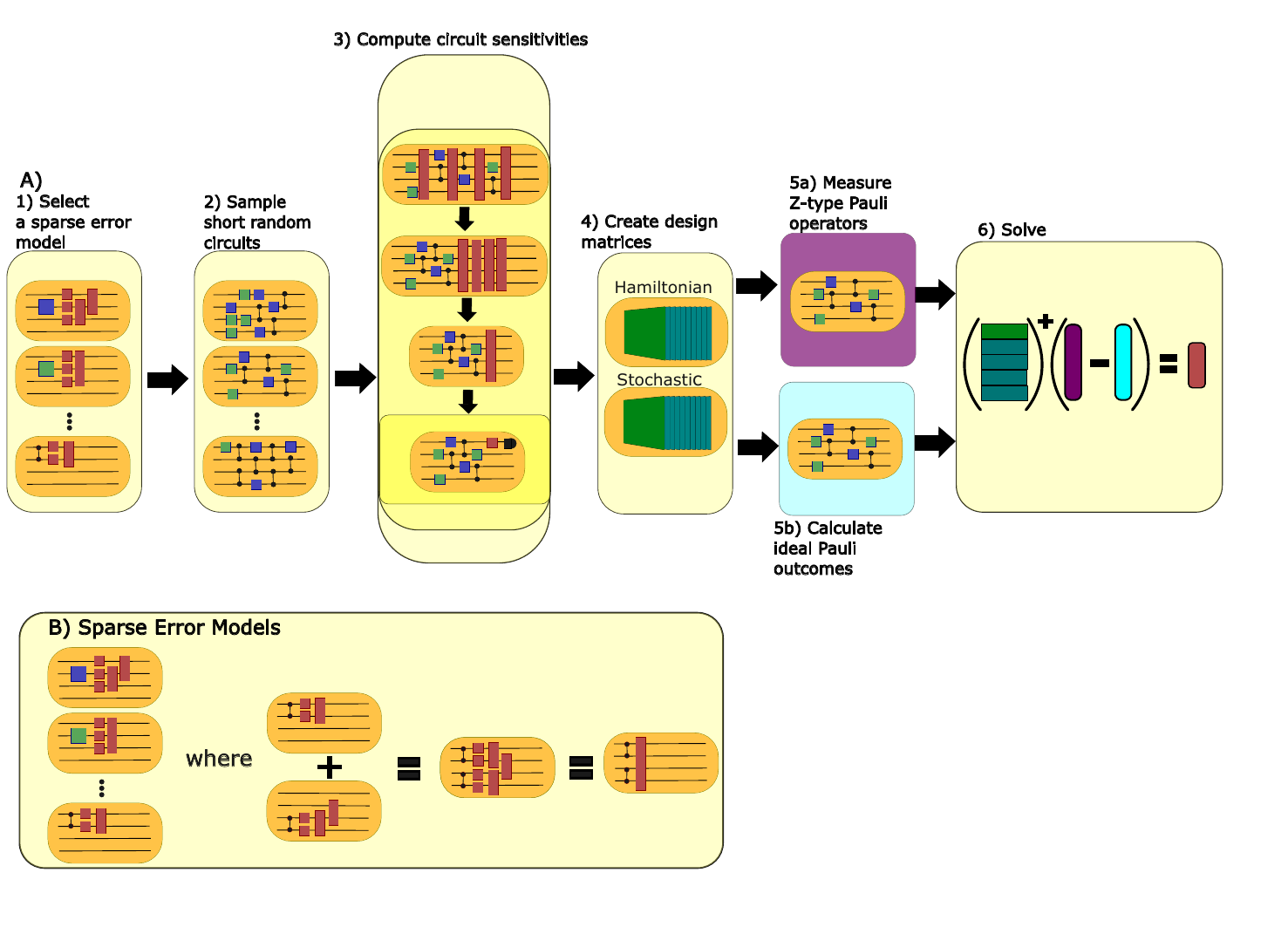}
    \caption{\textbf{Linearized GST.} An overview of linearized GST, which enables scalable learning of sparse error models. \textbf{A.} An overview of the procedure of linearized GST, which involves (1) selecting a sparse error model, (2) sampling short circuits to run, (3) computing the linear sensitivities of various $Z$-type Pauli observables to the model's parameters, (4) arranging those sensitivities into two design matrices (one for Hamiltonian and one for stochastic parameters), (5) estimating the Pauli observables for each circuit in experiment, and computing its error-free value, and then (6) solving two linear equations (one for Hamiltonian and one for stochastic parameters), by pseudo-inversion of the design matrix (for Hamiltonian parameters) and non-negative least squares (for stochastic parameters) \textbf{B.} The error models learned by linearized GST associate errors with gates and the error on a layer of gates as the sum of the Lindbladians associated with its constituent gates, as illustrated here.}
    \label{fig:algorithm}
\end{figure*}
\label{sec:SectionIII}
\subsection{Linearized sparse error models}\label{sec:error_models}
We now specify the structure of the error models that linearized GST is designed to learn. We consider an $n$ qubit system with a set of $n$-qubit layers $\{\ell\}$ constructed from some elementary operations (typically one- and two-qubit gates) $\{g\}$. It will simplify notation in this section to model SPAM error using dummy circuit layers (which we do not explicitly denote) that occur after initialization into the error-free $(\ket{0}\bra{0})^{\otimes n}$ state and prior to an error-free final measurement (which, in this paper, will always be a computational basis measurement that is processed to estimate the expectation values of one or more $Z$-type Pauli observables). 

Whereas standard GST \cite{gatesettomographyReview2021} and linear GST \cite{RLGST2021} model a layer's error by a dense $16^n \times 16^n$ process matrix, linearized GST models each layer's error by a \emph{sparse} error channel that can be efficiently described and learned.  We call an error channel sparse if it is well-approximated by the exponential of a Lindbladian that is a sum of at most $\textrm{poly}(n)$ of the $\mathcal{O}(16^n)$ elementary error generators -- i.e., if at most $\textrm{poly}(n)$ of its elementary error rates are non-zero. We additionally assume that a layer's error depends in a particular simple (linear) way on the identity of the gates that make it up.  Specifically, we associate to each \textit{gate} $g$ a list of errors with rates, and then the rate of elementary error generator $G$ in a layer $\ell$ comprising gates $g_1,\dots,g_k$ is given by $\epsilon_G = \epsilon_{G, g_1} +  \cdots + \epsilon_{G, g_k}$ where $\epsilon_{G,g}$ is the rate of error $G$ associated with gate $g$. The parameters of our error model are therefore the rates of $n$-qubit elementary error generators associated with elemental gates (typically one- and two-qubit gates) from which layers are constructed.  A gate $g$ may cause errors on any qubits, not just its targets, which allows us to model crosstalk easily.

We now state the structure of the error models used in linearized GST more precisely. In our $n$-qubit error models, a gate $g$ is associated with an $n$-qubit Lindbladian 
\begin{align}
\mathcal{L}_g =\sum_{P\in\mathbb{P}_{g,H}}h^{(g)}_{P}\mathbb{H}_{P}+\sum_{P\in\mathbb{P}_{g,S}}s^{(g)}_{P}\mathbb{S}_{P},
\end{align}
where $\mathbb{P}_{g,S}$ and $\mathbb{P}_{g,H}$ denote the subset of the non-identity $n$-qubit Pauli operators that indexes the set of Pauli stochastic and Hamiltonian errors, respectively, that, by assumption, can be nonzero for this gate. The sets $\mathbb{P}_{g,S}$ and $\mathbb{P}_{g,H}$ can contain Paulis that act on any qubits (not just the target[s] of $g$, but they must be of size at most $\textrm{poly}(n)$).  Their size determines the number of parameters that must be learned in linearized GST, and therefore the cost of performing the protocol. We model the error channel $\varepsilon_{\ell}$ associated with layer $\ell$ as: 
\begin{align}
\varepsilon_{\ell} &=\exp\left[\sum_{g\in\ell}\mathcal{L}_g\right]\\
&=\exp\left[\sum_{g\in \ell}\left( \sum_{P\in\mathbb{P}_{g,H}}h^{(g)}_{P}\mathbb{H}_{P}+\sum_{P\in\mathbb{P}_{g,S}}s^{(g)}_{P}\mathbb{S}_{P}\right)\right], 
\end{align}
where $\sum_{g \in \ell}$ denotes summing over all the gates in layer $\ell$. This defines the kind of error model that linearized GST is designed to learn. We arrange all of the unknown parameters in such an error model---i.e., all the $h_{P}^{(g)}$ and $s_{P}^{(g)}$---into a vector $\vec{\epsilon}$ using some arbitrary parameter ordering. Below, we show how to learn $\vec{\epsilon}$.

\subsection{Learning sparse error models with linear inversion}
We now show how learning the parameters of the above $n$-qubit error model, for an $n$-qubit layer set $\{\ell\}$, can be approximated as a linear inversion problem. Consider some circuit $C$ over the layer set $\{\ell\}$, and a Pauli observable $Q$ to be measured at the end of the circuit. We now calculate the effect of the model parameters on the Pauli observable $Q$. The superoperator for $C$ is given by
\begin{align}
    \tilde{\mathcal{U}} &= \prod_{\ell \in C} \varepsilon_{\ell} \mathcal{U}_{\ell}  \\ 
    &= \prod_{\ell\in C}\exp\left[\sum_{g\in \ell}\left( \sum_{P\in\mathbb{P}_{g,H}}h^{(g)}_{P}\mathbb{H}_{P}+\sum_{P\in\mathbb{P}_{g,S}}s^{(g)}_{P}\mathbb{S}_{P}\right)\right]\mathcal{U}_{\ell},
\end{align}
where $\mathcal{U}_{\ell}$ denotes the superoperator representation of the unitary associated with the error-free application of layer $\ell$. By applying the error propagation technique reviewed in Section~\ref{sec:SectionII} and summarized in Eqn. \eqref{Eqn:ErrorProp}, we can re-write the above equation as

\begin{align}
    \tilde{\mathcal{U}}= \prod_{\ell\in C}\exp\left[\sum_{g\in \ell}\left( \sum_{P\in\mathbb{P}_{g,H}} \gamma(U_{:\ell},P)h^{(g)}_{P}\mathbb{H}_{P'}+\sum_{P\in\mathbb{P}_{g,S}}s^{(g)}_{P}\mathbb{S}_{P'}\right)\right]\mathcal{U},
\end{align}
where $\mathcal{U}=\prod_{\ell \in C}\mathcal{U}_{\ell}$ is the ideal unitary superoperator enacted by the circuit, and $U_{:\ell}$ is product of all the unitaries for all the layers occurring after layer $\ell$. Applying the first-order approximation of Eqn.~\eqref{eq:circuit-first-order} we then obtain  
\begin{align}
     \tilde{\mathcal{U}}= \mathcal{U} + \sum_{\ell\in C}\sum_{g\in \ell}\left( \sum_{P\in\mathbb{P}_{g,H}} \gamma(U_{:\ell},P)h^{(g)}_{P}\mathbb{H}_{P'}+\sum_{P\in\mathbb{P}_{g,S}}s^{(g)}_{P}\mathbb{S}_{P'}\right)\mathcal{U},
     \label{eq:circuit-first-order-2}
\end{align}

We now calculate the effect of our model parameters---the $h_{P}^{(g)}$ and $s_{P}^{(g)}$---on a Pauli observable $Q$. We do so by applying Eqn.~\eqref{eqn:Ssen} and Eqn.~\eqref{eqn:Hsen} to Eqn.~\eqref{eq:circuit-first-order-2}. We express the result in terms of the \emph{change} in the Pauli observable 
\begin{equation}
     \Delta\langle Q\rangle_C =  \textrm{Tr}(Q\tilde{\mathcal{U}}[\ket{0}\bra{0}]) - \textrm{Tr}(Q\mathcal{U}[\ket{0}\bra{0}])
\end{equation}
Letting $\ket{\psi} = U \ket{0}^{\otimes n}$, we find that
\begin{align}
    \Delta\langle Q\rangle_C &=\sum_{\ell\in C}\sum_{g\in \ell} \sum_{P\in\mathbb{P}_{g,H}} \gamma(U_{:\ell},P)h^{(g)}_{P} \textrm{Tr}\left(Q\mathbb{H}_{P'}[\ket{\psi}\bra{\psi}]\right)
\end{align}
if the error-free expectation value (i.e., $\textrm{Tr}(Q[\ket{\psi}\bra{\psi}])$) is $0$, and 
\begin{align}
    \Delta\langle Q\rangle_C &=\sum_{\ell\in C}\sum_{g\in \ell} \sum_{P\in\mathbb{P}_{g,S}} s^{(g)}_{P} \textrm{Tr}\left(Q\mathbb{S}_{P'}[\ket{\psi}\bra{\psi}]\right)
\end{align}
otherwise (in which case it is -1 or 1). In either case, we can write
\begin{equation}
 \Delta\langle Q\rangle_C = \vec{s}_{Q,C}^T \vec{\epsilon}, \label{eq:linear_equation_1}
\end{equation}
where $\vec{\epsilon}$ is the vector of error model parameters, and $ \vec{s}_{Q,C}$ is the linear sensitivity to each parameter, which is computed directly from the above two equations---and which can be computed using only efficient-in-$n$ Pauli group and stabilizer state math. 

The analysis so far considers a single Pauli observable measured for a single circuit. We can, however, measure many Pauli observables for each circuit---from a computational basis measurement we can estimate any $Z$-type Pauli observable---and we can run many different circuits. Each such circuit $C$ and measurement $Q$ pair produces a different linear equation relating the observable change in $Q$'s expectation value $\Delta\langle Q\rangle_C$ to $\vec{\epsilon}$, as given by Eqn.~\eqref{eq:linear_equation_1}. Given $K$ circuit and measurement pairs, this can be summarized by:
\begin{equation}
\overrightarrow{\Delta\langle Q\rangle} =\mathbf{D} \vec{\epsilon}\label{eq:linear-relationship}
\end{equation}
where $\overrightarrow{\Delta\langle Q\rangle}$ is the length-$K$ vector of observed differences, $\vec{\epsilon}$ is the length-$\kappa$ vector of unknown error rates, and $\mathbf{D}$ is a $K \times \kappa$ matrix obtained by ``stacking'' the  $ \vec{s}_{Q,C}$.

Finding $\vec{\epsilon}$ amounts to solving the linear equation in Eqn.~\eqref{eq:linear-relationship}. If $\mathbf{D}$ is full rank (i.e., rank $\kappa$), we can solve for $\vec{\epsilon}$ using the Moore-Penrose pseudo-inverse:
\begin{equation}
 \vec{\epsilon} = \mathbf{D}^{+}\overrightarrow{\Delta\langle Q\rangle} \label{eq:linear-eq}
\end{equation}
where $M^{+}$ denote the pseudo-inverse of a matrix $M$. $\mathbf{D}$  will be full rank if enough circuits and measurements are chosen to obtain sensitivity to every parameter in the model, which is always possible unless there is a \emph{gauge freedom}  \cite{gatesettomographyReview2021} in the model. Gauge freedom in a model means that some model parameters or combinations of model parameters have no observable effect on any circuit and observable, which would prevent $\mathbf{D}$ from being full rank (in which case $\mathbf{D}$ can be inverted on its support). However, because linearized GST uses sparse error models there is not necessarily any gauge freedom---and in the particular error models used in our numerical examples we find that there is no gauge freedom (i.e. the design matrix is always full rank).

\begin{figure}
    \centering
    \includegraphics[width=\linewidth]{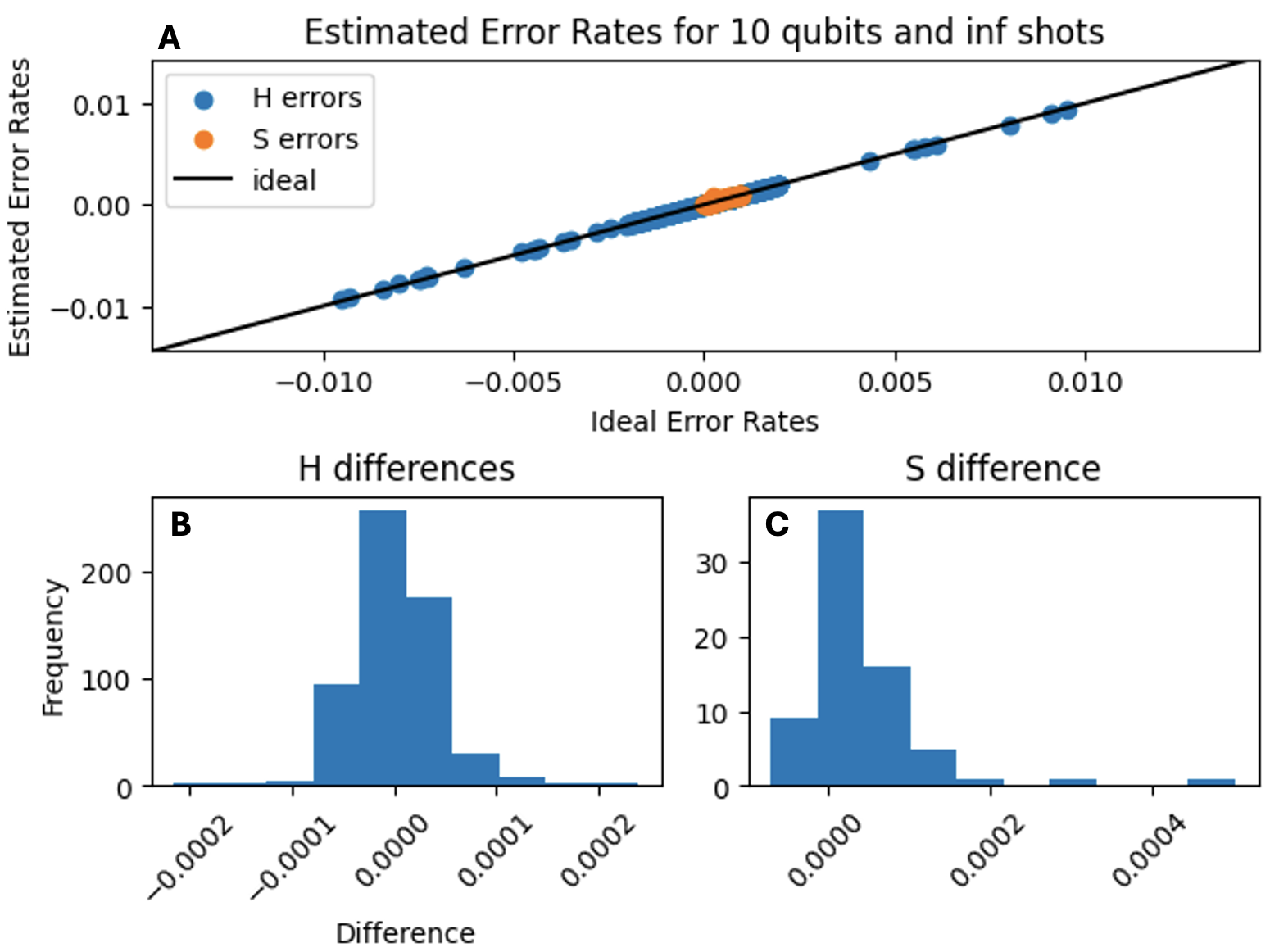}
    \includegraphics[width=\linewidth]{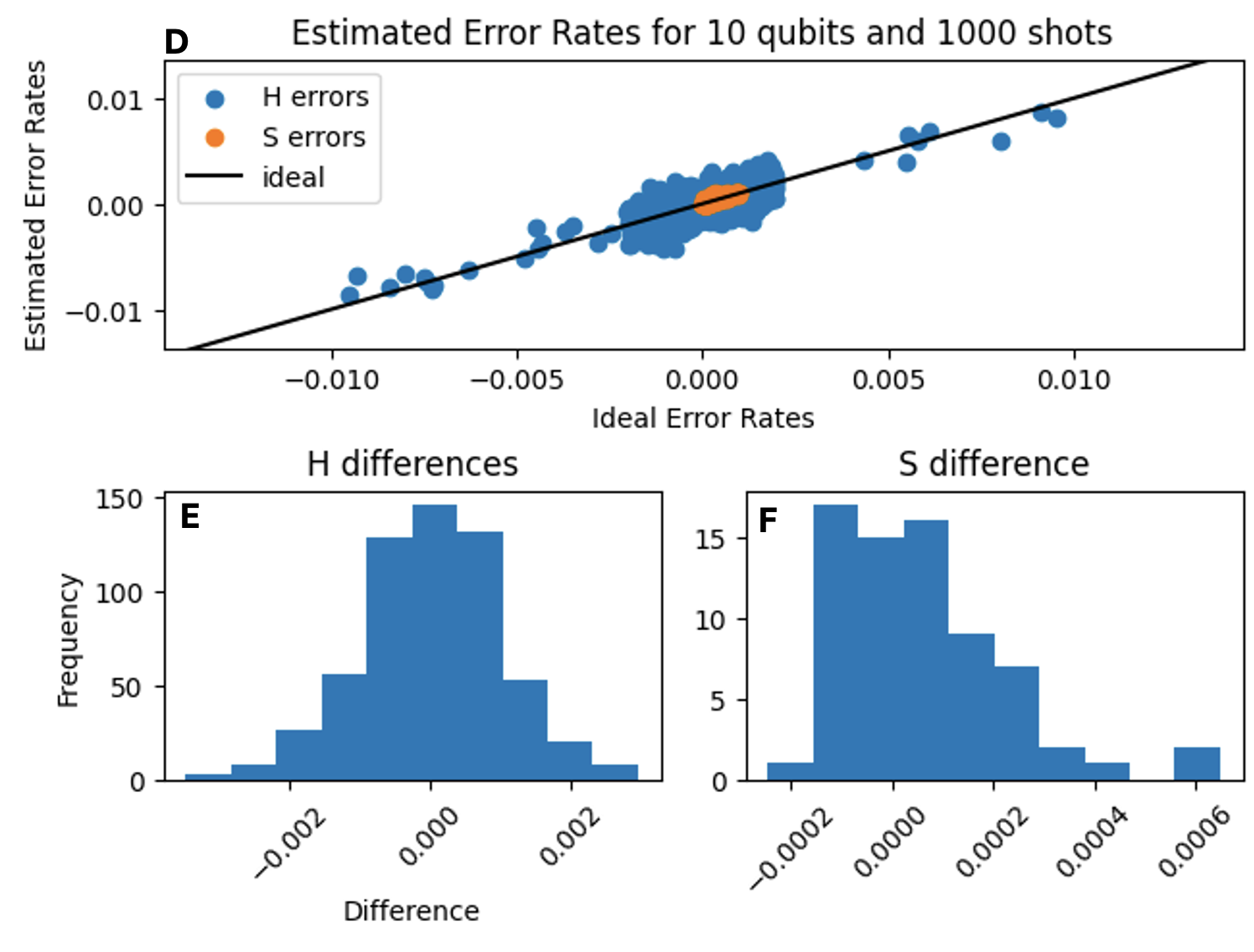}
        \caption{\textbf{Simulation of 10-qubit linearized GST.} The accuracy of linearized GST for a $10$-qubit error model containing local errors and crosstalk, using data from 1000 random circuits. \textbf{A, D.} The true error rates versus the estimate of linearized GST, separated into Hamiltonian (blue) and Pauli stochastic (orange) error rates, without shot noise (\textbf{A}) and with $N=1000$ shots for each circuit (\textbf{B}). The diagonal line represents no estimation error. \textbf{B, C.} The distribution of the absolute differences of the estimated and true values of the error rates, for Hamiltonian and Pauli stochastic errors, respectively, without shot noise. \textbf{E, F.} The distribution of the absolute differences of the estimated and true values of the error rates, for Hamiltonian and Pauli stochastic errors, respectively, with $N=1000$ shots per circuit. These results demonstrate that linearized GST can accurately learn  both coherent and stochastic errors in a 10-qubit system containing crosstalk.}
    \label{fig:simulation-1}
\end{figure}

Solving Eq.~\eqref{eq:linear-eq} to learn $\vec{\epsilon}$ is the central idea of linearized GST. In practice, however, we take a slightly different approach that enables enforcing the positivity constraint on stochastic error rates. Each circuit and Pauli observable pair is sensitive only to Hamiltonian or stochastic error parameters. This means that the Eqn.~\eqref{eq:linear-relationship} can be separated into two decoupled linear equations:
\begin{align}
\overrightarrow{\Delta\langle Q\rangle}_{\mathbb{H}} &=\mathbf{D}_{\mathbb{H}}\vec{\epsilon}_{\mathbb{H}}\label{eq:linear-relationship-H} \\
\overrightarrow{\Delta\langle Q\rangle}_{\mathbb{S}} &=\mathbf{D}_{\mathbb{S}}\vec{\epsilon}_{\mathbb{S}}\label{eq:linear-relationship-S},
\end{align}
where $\overrightarrow{\Delta\langle Q\rangle}_{\mathbb{H}}$ contains only those $C$ and $Q$ instances that are first-order sensitive to Hamiltonian parameters, $\mathbf{D}_{\mathbb{H}}$ is constructed by ``stacking'' the corresponding sensitivity vectors, and $\vec{\epsilon}_{\mathbb{H}}$ are the Hamiltonian parameters in the error model (and similarly for the $\mathbb{S}$-indexed equation). This is convenient for enforcing the complete positivity constraint on the error model. The stochastic error parameters must be positive, and we can enforce positivity by solving Eqn.~\eqref{eq:linear-relationship-S} subject to this constraint. This is known as the non-negative least squares problem:
\begin{align}
    \text{argmin}_{\vec{\epsilon}_{\mathbb{S}}} \left\lVert\mathbf{D}_{\mathbb{S}}\vec{\epsilon}_{\mathbb{S}}-\overrightarrow{\left\langle\Delta Q\right\rangle}_{\mathbb{S}}\right\rVert_2^2 \text{ subject to } \vec{\epsilon}_{\mathbb{S}}\geq0,\label{eq:S-solution}
\end{align}
and it is a convex optimization problem. In contrast, the Hamiltonian parameters can take any real value, and so we solve Eqn.~\eqref{eq:linear-relationship-H} using the pseudo-inverse:
\begin{align}
\vec{\epsilon}_{\mathbb{H}} =\mathbf{D}_{\mathbb{H}}^{+} \overrightarrow{\Delta\langle Q\rangle}_{\mathbb{H}}. \label{eq:H-solution}
\end{align}

The above derivation uses the exact Pauli expectation value differences. In practice, the differences in the Pauli expectation values $\overrightarrow{\Delta\langle Q\rangle}$ are estimated from a finite number of repetitions $N$ of each circuit, resulting in $O(1/\sqrt{N})$ ``shot noise'' uncertainties on each estimated Pauli expectation value (with correlations in the uncertainties for different Pauli observables computed for the same circuit). Uncertainties on the estimates due to shot noise can be easily computed using the linear propagation of error formula or a non-parametric bootstrap.

\subsection{The linearized GST protocol}
We now summarize the linearized GST protocol, which is a protocol for learning the parameters of a given error model with the structure specified in Section~\ref{sec:error_models}. This description assumes no gauge freedom in the error model. Linearized GST constructs a linearized approximation of the relationship between changes in Pauli expectation values and error model parameters $\vec{\epsilon}$, and uses it to estimate $\vec{\epsilon}$ by measuring Pauli expectation values for some set of circuits and then solving for $\vec{\epsilon}$ using Eqns.~\eqref{eq:H-solution} and ~\eqref{eq:S-solution}. This requires a procedure for picking circuits and Pauli measurements. If our model has no gauge freedom, then linearized GST requires a set of circuits and measurements that: (i) produce a full rank design matrix $\mathbf{D}$, and (ii) are all sufficiently \emph{shallow} that the linear approximation holds \footnote{The appropriate maximum depth depends on the magnitude of the errors, which is typically \emph{a priori} known approximately.}.

In this paper, we choose a set of $K$ random, shallow circuits that terminate in a computational basis measurement.  For each circuit, we estimate expectation values of all $Z$-type Pauli operators of weight up to some maximum $w$ using data from the terminating computational basis measurement. An elegant way to choose $K$ is to add circuits until $\mathbf{D}$ is full rank (or meets some criteria on its condition number), but in our simulations we simply set $K$ to constant and then confirm that the produced $\mathbf{D}$ is full rank.

\section{Simulations of Linearized GST}
\label{sec:SectionIV}
We now demonstrate linearized GST in simulation. We simulated a ten-qubit system and a five-qubit system with ring connectivity and a gate set consisting of $\frac{\pi}{2}$-rotations around the $X$, $Y$, and $Z$ axes on each qubit and $CZ$ two-qubit gates between connected qubits. All results below use simulations from the ten-qubit system unless otherwise stated. We simulated error models in which (i) each gate causes coherent and stochastic errors on its target qubit[s] that commute with the gate's ideal action, (ii) each gate causes local (single-qubit) coherent $Z$-rotation errors on all other qubits, and (iii) $CZ$ gates produce entangling coherent $ZZ$-rotation errors between each connected pair of qubit.  Both (ii) and (iii) are forms of crosstalk, with (ii) representing pulse spillover and (iii) representing unwanted coupling.  The error model also associates stochastic bit-flip ($S_X$) errors with state preparation and measurement (SPAM) operations, which tests and demonstrates linearized GST's ability to deal with SPAM errors. The rates of the errors in our simulated model are drawn at random from uniform distributions over intervals $[0,10^{-3}]$ (for stochastic errors) and $[-10^{-2},10^{-2}]$ (for coherent errors).  We constructed an experiment design by sampling 1000 depth-15 random circuits. For each of these circuits, we computed the expectation value (with no shot noise, representing the limit of $N\to\infty$ shots) of all weight-one and weight-two Z-type Pauli operators under this error model. All the results use data from these 1000 circuits, with shot noise added in some cases (as specified) and results from smaller datasets (fewer circuits) obtained by sub-sampling from this dataset.

\subsection{Linearized GST using the correct model ansatz}
We begin by demonstrating that linearized GST makes accurate predictions for the error rates both with and without shot noise (i.e., finite $N$ and $N\to \infty$ respectively) on the Pauli observables. An important component of linearized GST is the selection of the error model to be fit. In this first example we choose a parameterized model whose structure is identical to the data-generating model, but whose error \textit{rates} are unknown.  Figure~\ref{fig:simulation-1} shows the results of applying the linearized GST inversion algorithm to the above-described dataset both without shot noise ($N=\infty$) and with $N=1000$ shots per circuit. Figure~\ref{fig:simulation-1}A and D shows the true error rates versus the estimated error rates, obtained by applying GST to the dataset with and without shot noise, respectively. The estimated error rates are very close to the ideal error rates in the absence of shot noise---which quantifies the systematic error in linearized GST, in this example, due to the linear approximation---and the estimation error is still reasonable with $N=1000$. The estimation error is quantified in the histograms of Fig \ref{fig:simulation-1}B, C, E, and F, which show the differences between the true and estimated and ideal error rates for Hamiltonian and stochastic parameters without shot noise, and Hamiltonian and stochastic parameters with $N=1000$, respectively.

\begin{figure}
    \centering
    \includegraphics[width=\linewidth]{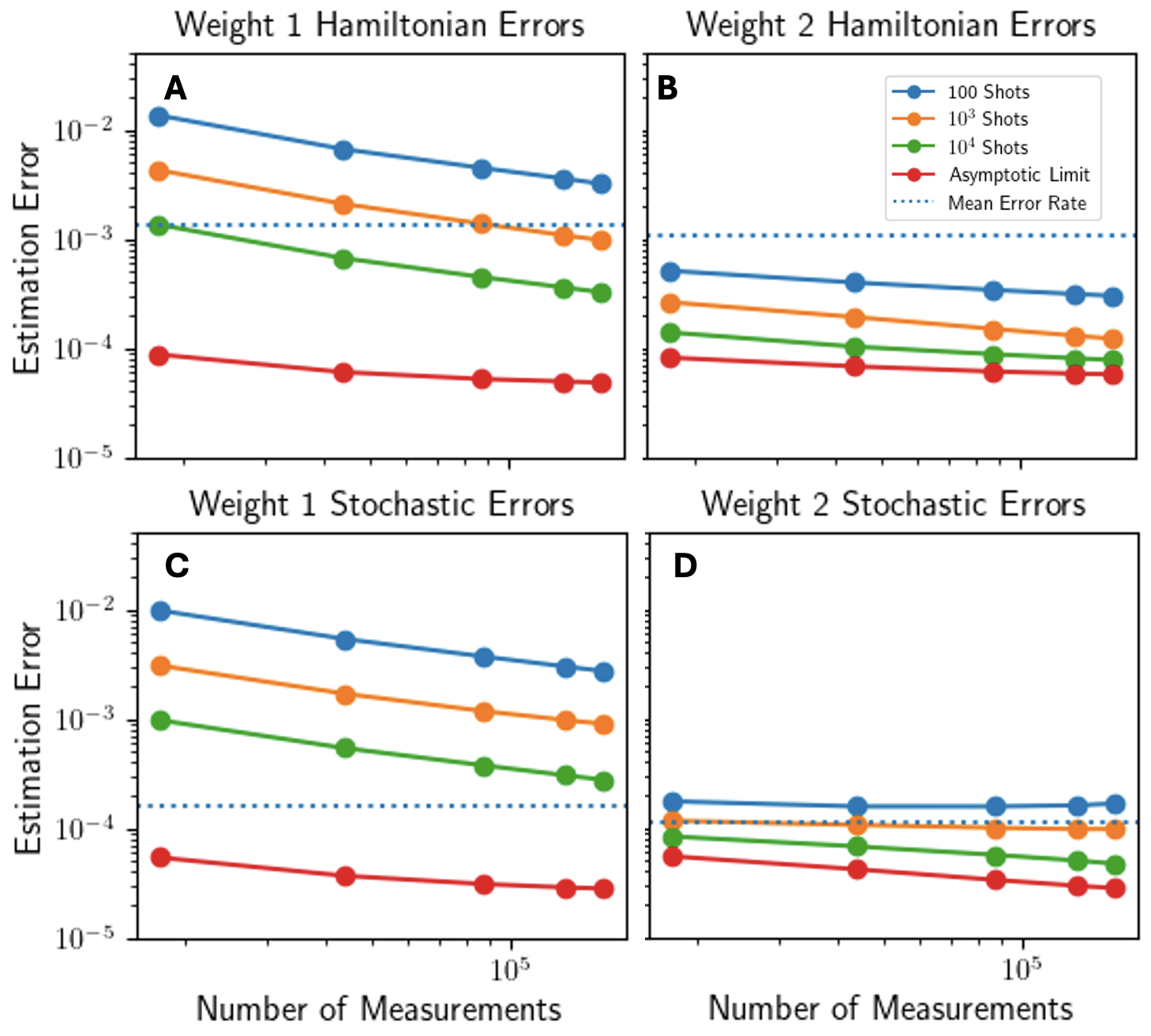}
    \caption{\textbf{Estimation error versus dataset size.} The average absolute error of the estimated errors as a function of the number of measurements performed (the number of circuits $K$ times the number of Pauli operators measured per circuit, which here is 55), divided into four kinds of error parameter in our error model: \textbf{A.} weight-one Hamiltonian errors, \textbf{B.} weight-two Hamiltonian errors, \textbf{C.} weight-one Pauli stochastic errors, and \textbf{D.} weight-two Pauli stochastic errors. The dotted line on each plot represents the average error magnitude of each error type. In each panel, we show the estimation errors for datasets with $N=100$ (blue points) $N=1000$ (orange points), $N=10000$ (green points), and $N=\infty$ (red points) shots per circuit.}
    \label{fig:CircuitShotNum}
\end{figure}

To explore the experimental cost of linearized GST we quantified how the accuracy of the learned error parameters scales in the number of total measurements performed ($K \times 55$, i.e., the number of circuits $K$ times the number of Pauli measurements, which here is $55$) and the number of shots for each circuit $N$. We do so by sub-sampling from our 1000-circuit dataset and adding varying amount of shot noise (varying $N$). Figure \ref{fig:CircuitShotNum} summarizes the estimation error of linearized GST versus the number of circuits for $N=100$, $N=1000$, $N=10000$ and $N = \infty$. Each data point in Fig. \ref{fig:CircuitShotNum} is produced by sub-sampling 500 different subsets of circuits of the given subset size and averaging the prediction error. Each point shows the average absolute difference between estimated error and the ideal error, for one of four classes of error parameters. Figures \ref{fig:CircuitShotNum}A-D summarize how accurately linearized GST estimated the rates of (A) weight-one Hamiltonian, (B) weight-two Hamiltonian, (C) weight-one Pauli stochastic, and (D) weight-two Pauli stochastic errors. We distinguish these four categories of estimation error because, in the simulated experiment, each category of errors has a different typical rate magnitude (indicated by the dotted line in each plot). Interestingly, we find lower estimation error for weight-2 Hamiltonian and stochastic parameters. In all cases, we find low absolute estimation error when $N$ is large.

\subsection{Linearized GST using the incorrect error model ansatz}
\begin{figure}
    \centering
    \includegraphics[width=\linewidth]{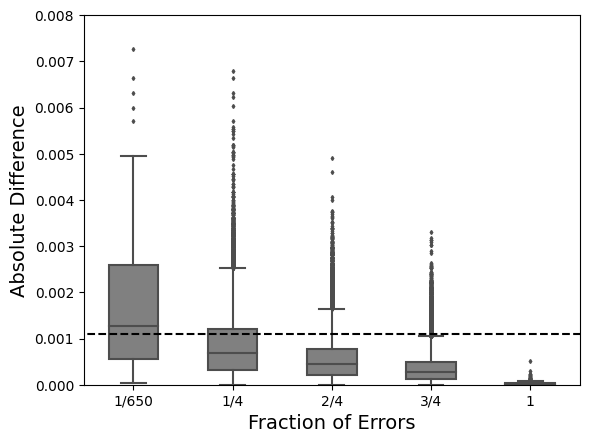}
    \caption{\textbf{Robustness of linearized GST to unmodeled errors.} The statistics of the absolute differences of the true error rates and the estimated error rates for linearized GST models that contain only some fraction $\eta$ of the parameters in the error-generator model. Each box plot shows the total differences across one hundred and fifty different error models with a randomly-sampled fraction $\eta$ of the error parameters, versus $\eta$. The dashed line shows the mean value of the true error rates in the data-generating model.}
    \label{fig:ReducedErrorModel}
\end{figure}

Linearized GST fits the parameters of an error model that must be chosen \emph{a priori}\footnote{although we expect compressed sensing techniques can be used to extend to \textit{a priori}-unknown sparse models}. In real-world use, users are unlikely to choose an ansatz that include exactly those errors whose rates really are non-negligible!  We therefore explored the robustness of linearized GST to choosing the wrong ansatz. In particular, we explore examples in which some errors have nonzero rates but are \textit{not} included in the fit ansatz.  To do so, we analyze the same data described above, but fit smaller error models (with some parameters removed) to it. We quantify the estimation error of linearized GST by the difference between error rates estimated by linearized GST and the true values for those error rates. 

We choose random incorrect ans\"atze models that contain only a fraction $\eta$ of the parameters in the data-generating model. The data-generating model contains $\kappa = 650$ parameters, so the smallest possible non-trivial value for $\eta$ is $\eta=1/\kappa = 1/650$, which yields an ansatz model with just one parameter.  We examine this case, and also $\eta=1/4$, $\eta=1/2$, $\eta=3/4$ and of course $\eta=1$. For each value of $\eta\neq 1$ we create and fit 150 different error models, each comprising $\eta \kappa$ of the error-generating model's parameters chosen at random.  We estimate the $\eta \kappa$ parameters of each ansatz using linearized GST. 

Figure \ref{fig:ReducedErrorModel} shows the distribution in estimation error, over parameters in each model and over the 100 models, versus $\eta$. For extreme (and unrealistic) case of $\eta=1/\kappa$ the estimation error is large. As $\eta \to 1$ the mean estimation error decreases, which is expected as the existence of unmodeled errors will typically pollute the estimates of modeled errors. In all cases except the single error case, the median error is below the magnitude of the mean error in the error-generating model, indicating estimates that are better than the error-free model. For $\eta = 3/4$, the estimation error is still significantly worse that obtained when we use the correct parameterized model ($\eta = 1$). This shows that linearized GST has some robustness to the existence of unmodeled errors, but that unmodeled errors do significantly pollute the parameter estimates.

\subsection{Effects of higher-order terms}
\begin{figure}
    \centering
    \includegraphics[width=\columnwidth]{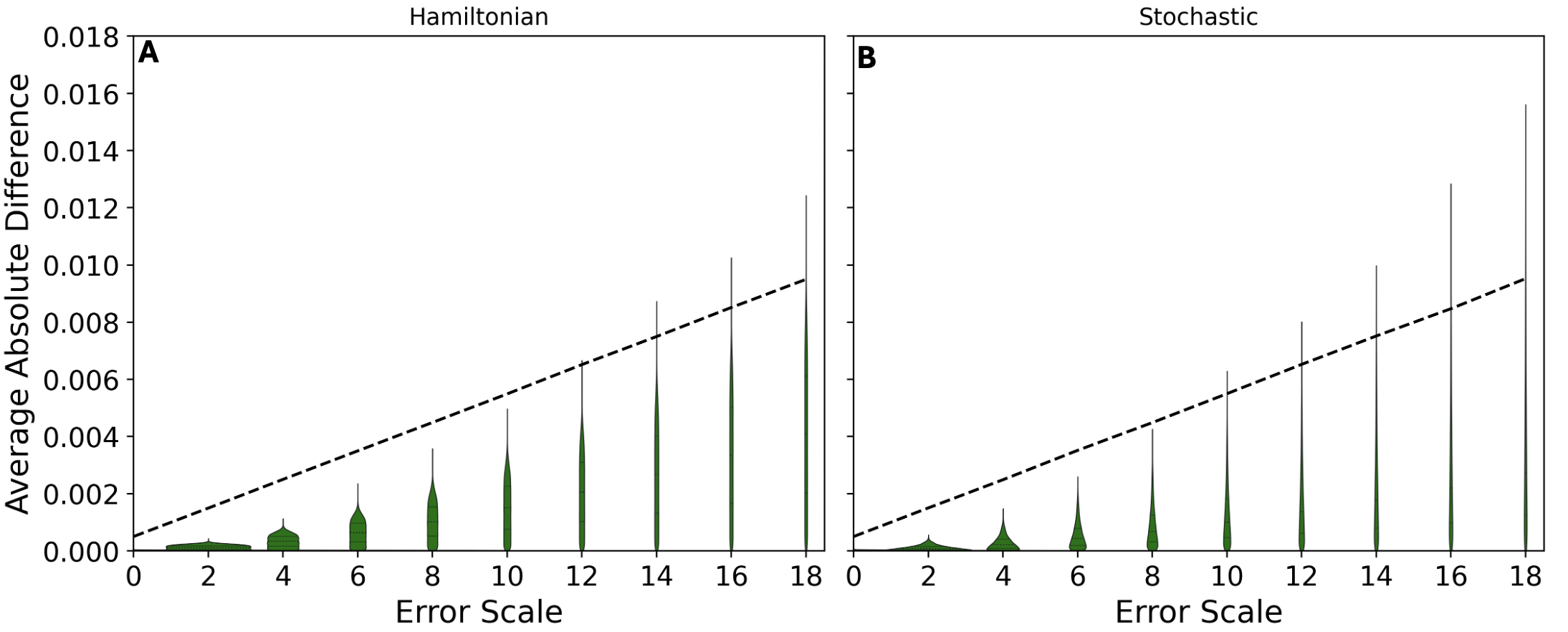}
    \caption{\textbf{Breakdown of the linear approximation.} Linearized GST relies on a linear approximation that only has low approximation error when the total magnitude of the errors in the circuits used in the linearized GST protocol is small. Here we demonstrate the breakdown in the linearized GST estimate's accuracy as the magnitude of the errors increase, in 5-qubit simulations. To do so, we scaled up the magnitude of the errors in an error model family, by a multiplicative ``error scale'' factor (the variable $c$ in the main text). Here we show the absolute error in the linearized GST estimates of each error parameter, over many different error models of each error magnitude, separated into the Hamiltonian and stochastic terms. The dashed line shows the mean absolute value of the error rates in the data-generating error models.}
    \label{fig:SystematicError}
\end{figure}

Linearized GST relies on a linear approximation that only has low approximation error when the total magnitude of the errors in the circuits used in the linearized GST protocol is small. To explore the  impact of the breakdown of this linear approximation on linearized GST's estimation accuracy, we simulated linearized GST on a five-qubit system with varying magnitude of error. In particular, we sample the error rates of the error-generating model as described above but multiply them by a factor of $c \geq 1$. We linearly vary $c$ between $c=1$ and $c=18$. For each value of $c$ we simulate the same set of 1000 circuits under 50 different randomly-sampled error models, producing 50 linearized GST estimates. For each dataset, we calculate the absolute difference between the estimated error rates and their true values.

Figure \ref{fig:SystematicError} shows the distribution in this absolute difference, for each value of $c$. We see that for small values of $c$, the median estimation error is much less than the average error rate (the dashed line), so the relative error in the estimate of a parameter is typically low. However, the median estimation error grows with $c$, and for large $c$ ($c \geq 12$) this median is around half the magnitude of the errors in the error-generating model---corresponding to a relative error that is typically large. This demonstrates that linearized GST will not provide reliable estimates when outside the small errors regime.

\subsection{Gauge freedom}
\begin{figure}
    \centering
    \includegraphics[width=6.5cm]{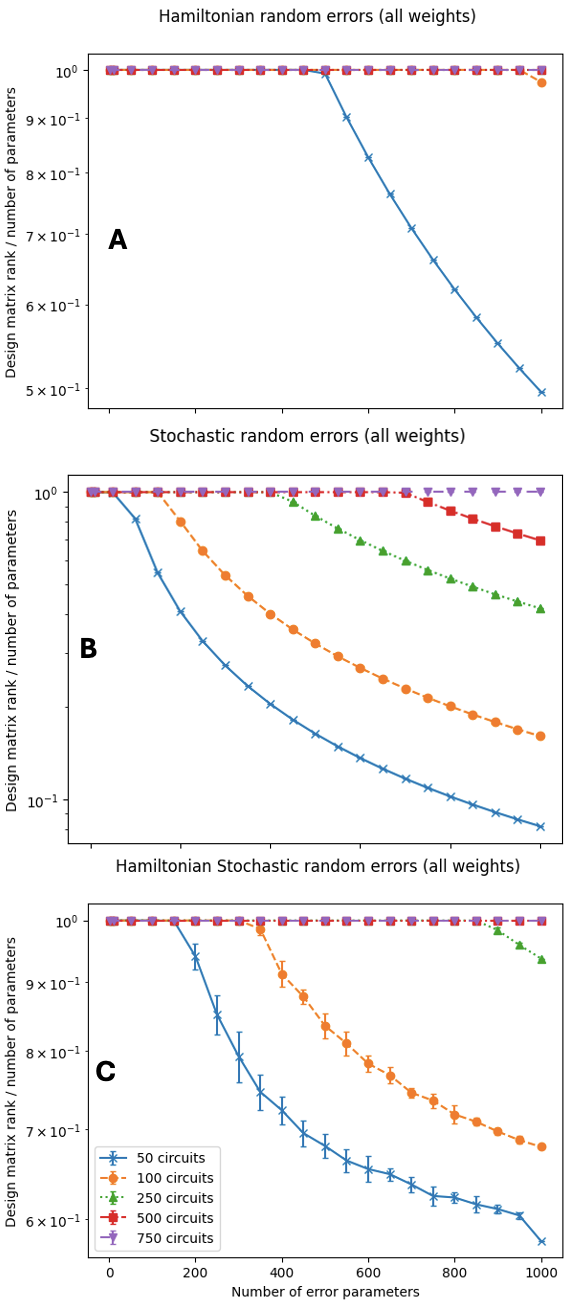}
    \caption{\textbf{Gauge freedom in sparse random error models.} We demonstrate that sparse error generator models do not typically have any gauge freedom, by showing that a full rank design matrix can be constructed for typical sparse error generator models with $\kappa$ parameters, as a function of $\kappa$. Here the sparse error generator models contain randomly-chosen elementary error generators as their parameters. We consider three cases: the elementary error generators are sampled from \textbf{A.} only the set of all Hamiltonian error generators, \textbf{B.} only the set of all stochastic error generators stochastic, and \textbf{C.} the of all Hamiltonian and stochastic error generators. We find that, with enough circuits, the rank of the design matrix divided by $\kappa$ is unity in every case, which implies no gauge freedom. We note, however, that more structured sparse error models are likely to have gauge freedom. }
    \label{fig:GaugeInvariance}
\end{figure}

The error models fit by linearized GST \textit{can} (but do not always) have gauge freedoms, like those that are present in the full error model of standard GST \cite{gatesettomographyReview2021}.   Whether a gauge freedom exists depends on the structure of the chosen error model ansatz. If there is a gauge freedom, a full rank design matrix $\mathbf{D}$ cannot be constructed, because the gauge parameters are unlearnable. In this case, an experiment design (circuits and Pauli observables) is sufficient if it enables learning all the \textit{non}-gauge parameters of the model.  To check this condition, we need to know the number of gauge parameters---but only if the error model has a gauge freedom.

We now explore whether gauge freedom typically appear in the error models used in linearized GST, where here we mean \emph{typically} in a statistical sense: i.e., when sampling parameterized error models from some distribution. We consider three classes of error model: those with only Hamiltonian error parameters, only stochastic error parameters, and a mixture of both in addition to SPAM error modeled by a stochastic $X$ error on each qubit. In each case, we pick a total of $\kappa$ $n$-qubit elementary error generators to include in the model, with each elementary error generator sampled uniformly from the set of all $n$-qubit elementary error generators of that type. For each instance, we sample $K$ circuits for various values of $K$, construct the design matrix $\mathbf{D}$, and compute its rank. 

Figure \ref{fig:GaugeInvariance} shows the rank of the design matrix divided by the number of parameters $\kappa$, which equals 1 iff $\mathbf{D}$ is full rank, as a function of $\kappa$. We find that in every randomly-sampled instance the design matrix is full rank for sufficiently many circuits. If a full-rank $\mathbf{D}$ can be constructed, then model has no gauge freedom.  These results therefore indicate that randomly-sampled models \textit{typically} have no gauge freedom. In practice, though, error models will not be randomly sampled. They will be constructed to reflect the anticipated errors in a system, and useful sparse error models may indeed have gauge freedoms. Finally, we note that Fig.~\ref{fig:GaugeInvariance} also offer interesting insights into the number of circuits typically needed to achieve full rank design matrices. For (at least) error models constructed with this random sampling procedure, a full-rank design matrix for Hamiltonian errors can be achieved using fewer circuits than are necessary to probe all stochastic errors.

\section{Discussion}\label{sec:conclusions}

Characterizing many-qubit systems in detail is historically difficult, because techniques like GST that enable detailed characterization require computational resources (time, memory) that scale exponentially with the number of qubits. The scalable characterization techniques available until now required (or imposed) very strong assumptions like strictly Pauli-stochastic errors (which can be enforced using Pauli frame randomization).  Such techniques cannot disambiguate stochastic and coherent errors, or quantify the rates of non-Pauli errors.  Our work, reported here, fills this gap with linearized GST, a scalable approach to GST that combines sparse error models, a linear approximation, and efficient simulation techniques to enable characterization of coherent and stochastic errors in many-qubit systems.

As presented here, linearized GST relies on an \emph{a priori} error model ansatz that only describes Hamiltonian (i.e., coherent) and Pauli-stochastic errors. We expect it will be straightforward to generalize linearized GST to characterize arbitrary small Markovian errors (by including active and Pauli-correlation error generators), and we believe this may be necessary for reliable, practical application of linearized GST due to the prevalence of amplitude damping errors (which are generated by a combination of stochastic Pauli and active elementary error generators) in quantum computing systems. We also expect it is possible to generalize linearized GST to scenarios in which a good error model ansatz is \textit{not} known \textit{a priori}, using compressed sensing or other model selection methods \cite{dobrynin2025compressed, PRXQuantum.6.020346}. The linearized GST data analysis involves solving a linear equation, so compressed sensing methods will be directly applicable \emph{if} a design matrix with appropriate properties (i.e., satisfying the restricted isometry property) can be constructed.

The linear approximation at the foundation of linearized GST limits the use of deep circuits to amplify errors and enable Heisenberg-like estimation precision \cite{gatesettomographyReview2021}. Errors -- especially coherent ones -- cannot be amplified beyond the point where 2nd order corrections become significant.  As a result, linearized GST will require more shots than traditional GST to achieve the same accuracy.  Achieving the same degree of Heisenberg-like precision as traditional GST will require a different approach. However, the approximate simulation methods on which linearized GST is built have a flexible approximation order, and so linearized GST could be generalized to an ``order-$k$ polynomial GST'' method where the data analysis would involve fitting an order-$k$ polynomial. Even $k=2$ might enable significantly deeper circuits and therefore more precise error rate estimates for a fixed number of shots per circuit.

The error models that we used here to demonstrate linearized GST did not have a gauge freedom, which simplifies the linearized GST procedure and interpretation of its results. Applying linearized GST to models with gauge freedoms is conceptually simple: the design matrix now needs to be full rank when projected onto the space of \emph{first-order gauge-invariant} properties of the model, and the linear inversion is performed on that subspace. This, however, requires scalable constructions of the first-order gauge-invariant subspace for a model. We anticipate that this also might be necessary for some uses of linearized GST.

Finally, linearized GST as presented here only characterized Clifford circuits, but it would remain tractable with many non-Clifford gates because we only require shallow circuits. However, since only Clifford gates are used in many error correction circuits, this does not prevent useful applications of linearized GST.

\section*{Acknowledgments}
This material was funded in part by the U.S. Department of Energy, Office of Science, Office of Advanced Scientific Computing Research, Quantum Testbed Pathfinder Program. T.P. acknowledges support from an Office of Advanced Scientific Computing Research Early Career Award. Sandia National Laboratories is a multi-program laboratory managed and operated by National Technology and Engineering Solutions of Sandia, LLC., a wholly owned subsidiary of Honeywell International, Inc., for the U.S. Department of Energy's National Nuclear Security Administration under contract DE-NA-0003525. All statements of fact, opinion or conclusions contained herein are those of the authors and should not be construed as representing the official views or policies of the U.S. Department of Energy or the U.S. Government.

\bibliography{main}
\end{document}